\documentstyle[preprint,aps,psfig]{revtex}

\newcommand{\be}{\begin{eqnarray}}
\newcommand{\ee}{\end{eqnarray}}

\begin{document}
\draft
\tighten

\title { Uncertainties in extracting $S_{17}$ from transfers}
\author{J.C. Fernandes\footnote{E-mail:jcff@wotan.ist.utl.pt},
R. Crespo\footnote{E-mail:raquel@wotan.ist.utl.pt}, 
F.M. Nunes\footnote{E-mail:filomena@wotan.ist.utl.pt}  }
\address{ Departamento de F\'{\i}sica, Instituto Superior T\'ecnico, \\
and Centro Multidisciplinar de Astrof\'{\i}sica (CENTRA) \\
Av Rovisco Pais 1096  Lisboa Codex, Portugal}
\author{I.J. Thompson\footnote{E-mail:I.Thompson@surrey.ac.uk }}
\address{ Department of Physics, University of Surrey, GU2 5XH, U.K.}

\date{\today}

\maketitle

\begin{abstract}

The transfer reaction  $^7{\rm Be(d,n)}^8{\rm B}$ is modelled
for the recent low experimental energies using the DWBA.
Tests of the validity of the ANC method for the extraction of the
$S_{17}$ astrophysical factor are performed.
We show that a peripheral assumption is no longer accurate
for center of mass energies greater than approximately 15  MeV.
In all cases we found that core deformation and/or excitation
effects to be small as long as the optical potential for the 
entrance channel is not strongly surface peaked.
Multistep corrections to the DWBA  were
found to be small.
We also emphasize  that the lack of knowledge of the optical potentials
for the entrance and outgoing channels induces severe uncertainties in the
extracted $S_{17}$ astrophysical factor.

\end{abstract}



\section{Introduction}

The proton rich nucleus $^8$B has been subject to very detailed studies
which are of relevance for the understanding of the  
so called solar neutrino problem \cite{seattle}.
Even if a nuclear solution cannot solve the
present disagreement between the  solar neutrino experiments, 
the $S_{17}$ astrophysical factor for the capture reaction
$^7$Be(p,$\gamma$)$^8$B
needs to be known with greater accuracy to established limits 
for the nonstandard theories \cite{crespo}.

Since the direct measurement of this reaction at the solar energies
is not possible due to the hindrance  by the Coulomb barrier, the 
measurements performed at higher energies are extrapolated 
(using some theoretical model) to the relevant
energy regime. As published data at higher energies differ considerably from
each other \cite{Kavanagh,Vaughn,Filippone,Parker,Kavanagh2,Wiez}, 
the corresponding low energy
reaction cross sections are rather ambiguous.
To remove this uncertainty 
new measurements of the capture cross section (at the lowest possible energies)
are being performed \cite{gilles}.

At the same time,
the structure of the light nuclei involved in these capture reactions is
incompletely known. The low energy behaviour of the $S_{17}$ is fairly
well established now, in particular after core degrees of freedom were
shown to have a negligible effect \cite{nunes1,nunes}. 
In contrast, the extracted  normalization of the S-factor is still poorly known.
It is strongly dependent on the size of 
the $^8$B nucleus and other observables \cite{csoto}.
If, on one hand, single particle models are suitable for predicting the 
low energy behaviour of the direct capture process, on the other hand 
they are unable to provide the normalization constant.
Because of the disagreement between data sets, there is no unique
determination of this normalization constant empirically from direct capture
measurements.

Recently the transfer reaction has been put forward as 
a new tool to extract the absolute value of the $S_{17}$ factor \cite{xu} as
long as the reaction is peripheral. Our aim here is to  check the
validity of this method.

The transfer reaction  $^7{\rm Be(d,n)}^8{\rm B}$ has been measured \cite{liu} 
at E$_{\rm cm}$=5.8 MeV and analyzed using the DWBA zero range 
approximation and a simple single particle structure model. 
Recently new data at  E$_{\rm cm}$=38.9 MeV from MSU has become available for
the same reaction \cite{msu}.
The measurement of the same reaction 
at the centre of mass energy of 15.6 MeV at the RIKEN Ring Cyclotron has 
also been performed \cite{Riken}.

In this work we analyze the transfer reaction  $^7{\rm Be(d,n)}^8{\rm B}$
for the relevant experimental energies using the DWBA  \cite{FRESCO}.
Special attention will be paid to the peripheral 
assumption. The uncertainties due to the lack of knowledge of
the optical potentials for the entrance and outgoing channels will
be studied. The contribution of the multistep corrections to the DWBA
arising from the low lying excited states in $^7$Be will be evaluated.

\section{Extracting the $S_{17}$ factor from transfers}

It is well established now that the direct capture S-factor 
$S^{\rm cap}_{17}$  derived theoretically
from  nuclear structure calculations can vary considerably,
and that although the energy dependence is fairly 
well known \cite{nunes,shoppa}, the
predicted astrophysical factors differ in their normalization \cite{Barker}.
Transfer reaction measurements, such as $^7$Be(d,n)$^8$B
are now being used as an indirect method
to extract an $S_{17}$ factor \cite{xu,liu} that should not be so model dependent.

The differential cross section for the  transfer reaction  
has been analyzed in 
ref. \cite{liu} using the DWBA. 
In this approximation it is assumed
that the $^7$Be core remains inert in the process, and the breakup component
on the deuteron wave function is only taken into account in so far that it affects
the optical potential.
The differential cross section is given by
\be
\frac{d \sigma}{d \Omega} = 
\frac{  \mu_{\rm i } \, \mu_{\rm f}}
{4 \pi^2 \hbar^4}
\frac{k_n}{k_d} \frac{1}{(2J_7 + 1) (2 J_d + 1) } \sum |T_{fi}|^2
~~,
\ee
with $ \mu_{ \rm i } $, \,
 $ \mu_{  \rm f }$ the reduced 
masses for the initial ($^7{\rm Be} - d$) and final ($^8{\rm B} - n$) 
channels and ${k_d},  \,   {k_n}$  the 
incident and outgoing momenta in the centre-of-mass frame. The transition 
amplitude for the transfer reaction process  in the prior form is
\be
T_{fi} = \sum  \langle \Psi_f^{(-)} 
 {\cal I}_{  \scriptsize  ^7{\rm Be} ^8{\rm B}} | V_{np}
 + V_{n^7{\rm Be}} - U_{n^8{\rm B}} | 
{\cal I}_{  \scriptsize  dn} \Psi_i^{(+)} \rangle ~~.
\label{TDWBA}
\ee
In this equation $\Psi_f^{(-)}$ and $ \Psi_i^{(+)}$ are the distorted waves
in the final and initial  channels respectively.
$ {\cal I}_{ \scriptsize  ^7{\rm Be} ^8{\rm B}}$ 
and ${\cal I}_{ \scriptsize  dn}$ are the internal
overlap wave functions for $^8$B (p - $^7$Be) and deuteron (p - n).
The remnant term, $V_{n^7{\rm Be}} - U_{n^8{\rm B}}$, where 
$V_{n^7{\rm Be}}$ is the interaction between the outgoing
neutron and the $^7$Be target, 
is usually small and can be neglected.
This term is included in our calculations but found to be small.  

Within  an uncorrelated  two-body model for the  p - $^7$Be  system,
 the proton occupies a single particle state  ${n \ell j}$
(p$_{3/2}$) and  the overlap function 
$ {\cal I}_{^7{\rm Be} ^8{\rm B}}$ can be defined as
\be  \label{scaling}
{\cal I}_{^7{\rm Be} ^8{\rm B}}(r) = {\cal S}^{1/2}   u_{n \ell j}(r)
\ee
where ${\cal S}$ is the spectroscopic factor that takes into account
the incompleteness of the two-body model in the description of the
$^8$B system and $ u_{n \ell j}(r)$ is the normalized single particle radial 
function.
Outside the range $R_N$ of the
$^7{\rm Be}$-p nuclear interaction, the overlap function becomes
\be
{\cal I}_{^7{\rm Be} ^8{\rm B}} \approx  {\cal S}^{1/2} 
\, b_{n \ell j}
W_{\eta \ell+ {\scriptstyle\frac{1}{2}} }(2 \kappa r) ~~~~~~~~~
r \gg R_N \label{overlap}
\ee
where $W_{\eta \ell+ {\scriptstyle\frac{1}{2}} }(2 \kappa r)$
is the Whittaker function, $\eta = Z(^7{\rm Be}) e^2 \mu / \kappa$ 
the Sommerfeld parameter, $\kappa = \sqrt{2 \mu \epsilon}/\hbar$,
$\mu$ the reduced mass for the (p--$^7$Be) system,
$\epsilon = 0.137$ MeV  the proton binding energy in $^8$B \cite{ajzenberg}
and $b_{nlj}$ the asymptotic normalization of the single particle radial
wave function $ u_{n \ell j}(r)$.

The asymptotic normalization constant (ANC) for the overlap function
${\cal I}_{^7{\rm Be} ^8{\rm B}}$ is defined by
\be
C_{n \ell j} = {\cal S}^{1/2}  \, b_{n \ell j} ~~. \label{ANC}
\ee
The ANC defines the probability of finding $^8$B in the configuration
p + $^7$Be
at distances outside the range of the  nuclear interaction.

Following a set of calculations for different $^8$B 
2--body models, where the proton occupies a single particle state
p$_{3/2}$, Xu {\em et al.} \cite{xu} found that the ANC for the overlap 
function is related to the $S_{17}$ factor by
\be
S_{17} = \frac{C_{p_{3/2}}^2}{0.026} \label{xueq}~~,
\ee
a relation that is independent of the g.s. model for $^8$B.
Using the definition of the ANC eq.(\ref{ANC}), an astrophysical factor
$S_{17}^I$ can therefore be determined from a spectroscopic factor and
the normalization $b_{3/2}$ of the p--$^7$Be wave function.
The spectroscopic factor in eq.(\ref{overlap}), 
is now  obtained (by a $\chi^2$ fit) from the ratio between the data and
DWBA calculations in the forward angle  region, 
and is denoted here by ${\cal S}_{\rm exp}$. Therefore
\be
S_{17}^{I} = \frac{b_{p_{3/2}}^2}{0.026} {\cal S}_{\rm exp} \label{xueq2}~~.
\ee
This method  of extracting the $S_{17}$  from transfers using 
the ANC will be referred as the ANC method.

We can alternatively use the direct capture calculations 
for extraction of the S-factor, by renormalizing the  direct capture 
astrophysical factor $S_{17}^{\rm cap}$ obtained from $u_{n \ell j}(r)$ 
 by the experimental spectroscopic 
factor ${\cal S}_{\rm exp}$,
\be
S_{17}^{II} = S_{17}^{\rm cap} {\cal S}_{\rm exp} \label{useq}~~.
\ee

Both methods rely on at least two assumptions. 
They assume  first that the wave function peripheral to $^8$B scales in 
proportion to the interior component,
as we have written in eq.(\ref{scaling}). Secondly, they assume that 
measurements of forward-angle transfer cross sections are sufficient
to determine the spectroscopic factor.  
If DWBA analyses are used for this determination,
then it is assumed further  that the DWBA  gives
a good description for the transfer reaction, and that the distorted waves
for the incoming and outgoing channels are well determined. 

The ANC method of eq.(\ref{xueq2}) is only reliable when the
transfer reaction is peripheral. 
In such cases the asymptotic part
of the overlap function of eq.(\ref{overlap})
is the dominant contributor to the transition amplitude eq.(\ref{TDWBA}).
For peripheral transfer reactions the results obtained through both methods
are equivalent, while the method defined in eq.(\ref{useq}) can be used as a 
generalization of the ANC method when multistep effects are included in the 
reaction mechanism.

\section{The $^8$B ground state wave function}

Many two-body models have been developed for the description of $^8$B 
\cite{kim,tom,rob,esb} that assume a spherical $^7$Be  core. The 
large quadrupole moment of the mirror nucleus   $^7$Li 
suggests however that $^7$Be should be  considerably deformed.
The effects of core deformation on the capture reaction were studied in ref. 
\cite{nunes}
using a  core + proton model where the core is deformed and allowed to excite.
When the $^{7}$Be core is assumed to be deformed then the g.s. of $^8$B ($2^+$)
will have a contribution from the $p_{1/2}$ proton coupled to
the $^7$Be($\frac{3}{2}^-$) channel ($p_{1/2} \otimes \frac{3}{2}^-$),
as well as ($p_{3/2} \otimes \frac{3}{2}^-$). 
Within that model, if the deformation of the 
$^7$Be core is taken to be $\beta_2$ = +0.5  the probability of
finding the system in the channel ($p_{3/2} \otimes \frac{3}{2}^-$)
is significantly reduced from 1.0 (no deformation) to 0.57  ({\tt reo1}),
while for  $\beta_2$ = -0.5 ({\tt reo2})
it was found that $P[p_{3/2} \otimes \frac{3}{2} ] = 0.94$.

The $^{7}$Be has a low lying excited state  
E($\frac{1}{2}^-$) $= 0.43$ MeV and a higher spin excited state
at E($\frac{7}{2}^-$) $= 4.57$ MeV. 
Due to its low excitation energy, the $\frac{1}{2}^-$ state  
couples easily to the g.s. 
while the  $\frac{7}{2}^-$ state  also
strongly couples to the ground state.
Including core excitation when calculating the g.s. of $^{8}$B
enables two core excited channels that are important for
the g.s. of $^8$B: $(p_{3/2} \otimes \frac{1}{2}^-)$ and 
$(p_{3/2} \otimes \frac{7}{2}^-)$.
The inclusion of excitation modifies the $^8$B g.s., reducing
the normalization of the ($p_{3/2} \otimes \frac{3}{2}^-$) channel
as in the case of deformation. It was found that
$\beta_2$ = +0.5 ({\tt exc1})  gives
$P[p_{3/2} \otimes \frac{3}{2} ] = 0.39$, and $\beta_2$ = -0.5 
gives ({\tt exc2}) $P[p_{3/2} \otimes \frac{3}{2} ] = 0.87$.

These distinct models of $^8$B will be used to test the validity
of the ANC method, by seeing how well the extracted $S_{17}$ differs
from that of the one-channel models upon the introduction of
deformation/excitation and multistep core couplings.

\section{Results}

\subsection{Sensitivity to the peripheral assumption}

In order to analyze the uncertainties associated with the ANC method from
assuming a peripheral character for the transfer reaction, we show in
fig.(\ref{fig:xsecl}) the reaction cross section as a function of the 
partial wave and the corresponding differential cross section
for energies E$_{\rm cm}$ = 5.8 MeV, 15.6 MeV and  38.9 MeV.
The g.s. model for $^8$B was taken from Kim \cite{kim}.
For E$_{\rm cm}$ = 5.8 MeV we take the
optical potential parameters from d--$^9$Be elastic scattering at
E$_{\rm cm}$ = 6.4 MeV of Satchler  \cite{jc4}
and from n--$^9$Be at the correct energy of Dave and  Gould  \cite{jc8}
for the income and outgoing channel respectively.
For the RIKEN experiment at 13.6 MeV we use the potential parameters from
d--$^7$Li at  E$_{\rm cm}$ = 18.7 MeV of El-Nadi \cite{jc6} and
n--$^9$Be at   E$_{\rm cm}$= 19.44 MeV of Olsson {\em et al}. \cite{Olsson} 
respectively. 
Since in the case of the MSU experiment at 38.9 MeV the quasi-elastic scattering 
was
measured for the entrance channel, we take the optical potential
parameters for this channel, by extracting the elastic scattering cross section 
data
from the quasi-elastic data \cite{msu}.
For the outgoing channel we use the potential defined in ref. \cite{Olsson}.

The transition amplitude  eq.(\ref{TDWBA}) involves a radial integral
over the projectile-target relative distances of the
incoming and outgoing distorted waves. 
The contribution from the outgoing channel to
this integral was evaluated for radii from R$_{\rm cut}$ to infinity for
several values for the cutoff radius.
For E$_{\rm cm}$ = 5.8 MeV the cross section peaks around $L$ = 3 
corresponding to an impact parameter of 4.5 fm which is significantly
larger than the $^8$B interaction  radius of 2.95 fm \cite{surrey}
and thus, for this energy, the transfer is clearly peripheral. 
For the MSU energy, E$_{\rm cm}$ = 38.9 MeV,  the cross section peaks
for partial wave values  around
$L$ = 4 which corresponds to an impact
parameter of 2.3 fm and thus one can expect a significant
contribution from the nuclear interior to the transition amplitude radial 
integral. This is illustrated in fig.(\ref{fig:xsecl}) where the reaction 
cross section for MSU  becomes significantly smaller for a cutoff radius
R$_{\rm cut}$ of the order of the nuclear radius. This contrasts
with the lower energy case of
E$_{\rm cm}$ = 5.8 MeV, where the contribution for the reaction cross section
becomes negligible only for very large cutoff radius 
${\rm R}_{\rm cut} \gg {\rm R}_{\rm N}$.
It follows that, at the energies of the MSU experiment, the ANC method cannot 
be used by itself to extract the $S_{17}$ factor accurately.
For the RIKEN experiment, the impact parameter is 3.3 fm which is only
slightly larger than the $^8$B interaction  radius, thus caution should be
taken when extracting the $S_{17}$ from the transfer reaction at this energy.

\subsection{Sensitivity to the optical potential inputs}

The distorted waves for the incoming and outgoing channels are one
of the inputs of the DWBA. The lack of measurements for
the initial and final channel elastic scattering  induces uncertainties
in the $S_{17}$ extracted from the transfer studies,
uncertainties which are accentuated when one is using elastic scattering data 
either at a shifted energy or on a different target (or both).

A systematic study on the optical potential uncertainties was performed
recently \cite{comm,us} for the $^7$Be(d,n)$^8$B reaction at
E$_{\rm cm}$ = 5.8 MeV. We have shown in ref. \cite{us} that if one
uses different sets of optical potential parameters for the entrance channel
\cite{liu,jc2,jc3} that fit the elastic scattering of deuterons from 
$^7$Li at higher energies (E$_{\rm cm} \simeq 9.3 $ MeV) the calculated 
transfer cross section varies considerably, producing up to $16\%$ 
uncertainty in the extracted $S_{17}(0)$ (calculated relative to the
mean value). This  is illustrated in fig.(2) which shows that 
the calculated d--$^7$Be elastic scattering at the 
correct energy (E$_{\rm cm}$ = 5.8 MeV) using this set of potentials 
are different (fig.2a), even at small angles. Consequently, the calculated 
transfer cross section varies considerably (see fig. 2b).

Apart from the error introduced by extracting the optical potential from
elastic scattering at the wrong energy, there is also an error due to the
changes in the mass or charge of the target. Fig.(3) shows the calculated 
elastic differential cross section (fig. 3a) using deuteron
optical parameters from elastic scattering on $^6$Li and $^9$Be at the
correct energy \cite{jc4,jc3} and the corresponding transfer cross
section (fig.3b).
We find that the uncertainty on the $S_{17}(0)$ relative to its mean value is
$20\%$.
For both these calculations we have used a n--$^9$Be potential from \cite{jc8} 
for the exit channel and Kim's single particle potential from $^8$B \cite{kim}.

As for the outgoing channel, the combined energy and mass/charge
uncertainties in defining the optical potentials for the outgoing channel 
were found to be  of order of $12\%$.

If we take the whole range of optical potentials for entrance and exit channels  
mentioned in \cite{us}, using eq.(\ref{xueq}), one 
obtains for the S-factor of
$S_{17} = 26.2 \pm 8.4$ eV b. These
results show a theoretical uncertainty larger
than indicated in \cite{liu} and are consistent with the recent work by
the Texas A\&M group \cite{comm}. However,
one can minimize this uncertainty by using only potential parameters taken
from elastic scattering at the correct energy.
If we reduce our set to d-$^6$Li or d-$^9$Be optical potentials from
\cite{jc4,jc3} at the correct energy for the entrance channel
and n-$^9$Be optical potentials from \cite{jc8} at the correct energy for the
exit channel, we obtain $S_{17} = 23.5 \pm 3.7 $ eV b
which indicates a smaller prediction for
the S-factor than  the result  $S_{17}$ = 27.4 $\pm$ 4.4 eV b
presented in \cite{liu}.

The uncertainty on the extraction of $S_{17}$ from transfer reactions
induced by the lack of knowledge of the optical potentials for
the incoming and outgoing channels is also evident at higher energies,
such as, for example, for the RIKEN measurement.
Fig.(\ref{fig:Riken}a) shows  the calculated  elastic scattering
for the incoming channel using the optical potential parameters taken from 
d--$^7$Li but at different center of mass energies of El-Nadi {\em et al.} at
E$_{\rm cm}$ = 18.7 MeV \cite{jc6} (solid) and 
Matsuki {\em et al.} at 11.43 MeV \cite{jc1} (dashed).
Fig.(\ref{fig:Riken}b) shows the
corresponding transfer differential cross section. 
The  solid and the dashed curves were evaluated using the El-Nadi and 
Matsuki potentials respectively for the incoming channel and the potentials of
\cite{Olsson} for the outgoing channel.
The dotted solid and dotted dashed  lines were obtained  using the potentials
El-Nadi and Matsuki respectively for the incoming channel, but with the 
potentials defined in \cite{jc8} for the outgoing channel.  
Once more, the uncertainties on the incoming channel have large effects
on the calculated transfer reaction. The effect of the uncertainties in
the outgoing channel on the transfer is comparatively less important.

\subsection{Sensitivity to the $^8$B ground state wave function}

We now study the independence of the ANC method with
respect to the choice of different two-body
models for the description of the $^8$B ground state. In particular, 
we study the effect of including the $^7$Be core deformation and/or excitation.
To avoid uncertainties on the calculated transfer differential cross section 
by modifying  the incoming optical potential by distortion, 
we verified that  elastic scattering is 
not significantly altered by the inclusion of 
deformation and/or core excitation.
We show first, in fig.(\ref{fig:B8opt}),  the calculated differential
cross section at E$_{\rm cm}$ = 5.8 MeV using different single
particle 2 body--models with an inert core for the description of $^8$B.
The optical potential parameters for the outgoing channel were taken to be
those given in section A. 
The optical potential parameters for the incoming channel are taken from 
Bingham \cite{jc3} denoted here as the Surface I potential.
As follows from 
eqs.(\ref{ANC}-\ref{useq}), the ratio of the asymptotic normalization
of the single particle radial wave function to the calculated 
capture S-factor, $b^2/S_{17}^{\rm cap}$, should be model 
independent and equal to 0.026 (eV$^{-1}$ b$^{-1}$ fm$^{-1}$)
 for the case of one-channel single particle models.
Table \ref{as-ratio} shows that the calculated ratio
for the single particle models of Kim \cite{kim}, Tombrello \cite{tom},
Robertson \cite{rob}, Barker \cite{Barker}, and Esbensen \cite{esb} 
are constant, in agreement with Xu {\em et al.} \cite{xu}.
For the purpose of estimating
the effect of using different $^8$B g.s. models, we take the astrophysical
S-factor calculated for direct capture at 20 keV. 
In the table \ref{S17-factors} we compare the $S_{17}^I$ factor evaluated for 
different two body models using the ANC relation eq.(\ref{xueq2}) for
the data at E$_{cm}$=5.8 MeV, with the capture
normalization constant $S_{17}^{II}$ of eq.(\ref{useq}).
To a good accuracy, both methods consistently show that 
the ANC method can deduce the $S_{17}$ from the E$_{cm}$=5.8 MeV experiment
in a manner that is essentially independent of the description of the 
$^8$B ground state in the case of one-channel models. 
Taking Kim model as a reference, we can see 
that the extracted $S_{17}$ factors for different g.s. models
differ by less than 5$\%$.

When introducing deformation one could na\"{\i}vely expect no 
change in the relationship between the extracted S-factor and the
ANC of the $^8$B g.s. wave function.
The $^8$B g.s. wavefunction can be written as 
$|\Psi \rangle = \alpha_1 |\Psi_{p_{1/2}} \rangle ~ |^7{\rm Be _{g.s.}} \rangle
 +  \alpha_2 |\Psi_{p_{3/2}} \rangle  ~ |^7{\rm Be_{g.s.}} \rangle $
and as both radial functions $\Psi_{p_{1/2}}$ and  $\Psi_{p_{3/2}}$
have the same asymptotic behaviour, we have $b_{1/2} =  b_{3/2}$.
Given that the $^8$B wave function is normalized,
$\alpha_1^2 +  \alpha_2^2 = 1$. Therefore, whatever the deformation, the 
single particle normalization constant is the same: 
$b^2 = b_{1/2}^2 = b_{3/2}^2$.
When including excitation, since the contribution of the excited
states decay rapidly to zero their contribution to the $S_{17}$ factor
is expected to be negligible, as found in \cite{nunes}, and thus their 
contribution can be neglected when evaluating $b$ with eq.(\ref{xueq2}).
As shown in table \ref{as-ratio} the ANC eq.(\ref{xueq}) of Xu {\em et al.}  for 
extracting the $S_{17}$ remains valid.

In contrast to the results shown in Table \ref{as-ratio},
 we find that the DWBA cross section and
 therefore the S-factor 
are sensitive to the deformation and/or excitation for strong surface peaked
optical potentials in the entrance channel. This can be seen in 
Table \ref{S17-factors}, where we
compare the extracted S-factor using an optical potential for
the entrance channel \cite{jc4} that is strongly surface peaked,
(here denoted by Surface II), with the results obtained  using the 
Surface I optical potential.
The $\chi^2$ is defined according to
$ \chi^2 = \frac{1}{N_{\rm exp}}
 \sum_i \sqrt{
\left( \frac{ {S}_{\rm exp}(i) -  {S}_{\rm Theo}(i) }
{\Delta {S}_{\rm exp}(i)}             \right)^2
 }$ with $N_{\rm exp}$ the number of experimental points and 
$\Delta {S}_{\rm exp}(i)$ the experimental error associated with each point.
For no deformation or excitation, 
the extracted S-factor is essentially model independent for both optical
potentials. In contrast,  
the extracted $S_{17}$ with the {\tt reo1} model 
(which has a larger component for the 
$[p_{3/2}\otimes\frac{3}{2}]$ channel) differs from the Kim model  
by only 1.4 $\%$ for Surface I, but 16 $\%$ for Surface II.
The effects are similar for {\tt exc1} which has as well a larger 
component for the $[p_{3/2}\otimes\frac{3}{2}]$ channel.
This is due to the fact that deformation modifies the low partial waves
that are contributing significantly to the cross section. 
These modifications are probed by optical potentials which are strongly
surface peaked for the incoming channel and have less interior absorption.

\subsection{Multistep processes}

The DWBA formalism assumes that the inelasticity arising from core excitation
of the projectile and/or excitation of either projectile or target, 
has a small effect on the transfers. 
The low lying excited states of  $^7$Be could in principle be excited by the 
deuteron prior to the transfer, giving rise to multiple step processes. 
In order to estimate the
contribution of these processes to the differential cross section, we
consider that the core ($^7$Be) is deformed, and
consider quadrupole rotational nuclear couplings between its
3 bound states. 
In doing so, the transfer process will depend on couplings to and from
the g.s. of $^7$Be and the excited states, in addition to the transfer
couplings themselves.

A diagram of the allowed 1, 2 and 3-step paths, excluding
back transfer couplings, is shown in
fig.(\ref{fig:diag}).
The observables were calculated including all couplings between 
the 3 states of $^7$Be and transfer couplings between the 3 states of
$^7$Be and the g.s.  of $^8$B. We used Surface I optical potential
for the entrance  and the potential defined in \cite{jc8} for the exit channel.

Fig.(\ref{fig:multi}a) shows that the 2-step corrections to the
elastic cross section calculated with model {\tt exc1} from \cite{nunes})
are small, being evident only at large angles and that higher order 
multiple step corrections are small. 
Fig.(\ref{fig:multi}b) also shows that the transfer cross section is not 
strongly sensitive to the multistep effects. They reduce the differential
cross section peak by $\sim$ 8$\%$. 
As a direct consequence, the effect on the extracted $S_{17}$-factor (evaluated by renormalizing 
$S^{\rm cap}_{17}$ by the experimental spectroscopic factor 
extracted from the transfer reaction) is around $\sim$ 8$\%$.
This same result was confirmed using other
core deformation/excitation  models defined in \cite{nunes} and with
other optical potentials for the incoming and outgoing channel.

It is known that when introducing back couplings
to the original DWBA specification,
the optical potential fit to the scattering data may well be lost  
\cite{back}. In fig.(\ref{fig:coup}) we show the elastic scattering
fig.(\ref{fig:coup}a) and transfer cross section fig.(\ref{fig:coup}b)
for 3 cases:
a) all couplings between the core states are included but no back
transfer couplings ; b) all couplings 
including back transfer couplings; c) no back transfer couplings and only up
couplings for the core.
As shown in the figure, the effect of introducing back transfer couplings 
is of same order of magnitude as
the couplings from the excited states to the g.s. of $^7$Be.

The uncertainty arising from losing the elastic scattering fit, when 
introducing back transfer couplings, may be circumvented if we  assume that 
these couplings 
are already effectively included in the optical potentials, 
and, to avoid double counting, should therefore be removed from our 
transfer calculations.
This result calls once more for measuring  the elastic scattering 
channels if one wants to avoid evaluating theoretically the second
order coupling between the elastic and transfer 
channels.

\section{Conclusions}

We  analyzed the transfer reaction  $^7{\rm Be(d,n)}^8{\rm B}$ 
for the recent experimental energies using DWBA.
Checks on the ANC method for extracting the $S_{17}$ astrophysical
factor from the transfer reaction were performed.
We have shown that the lack of experimental data for the elastic scattering
incoming and outgoing channels induces severe uncertainties on the
extraction of the  astrophysical factor.
 From the low energy transfer data of \cite{liu} we obtain 
a range for the S-factor $S_{17}(0) = 23.5 \pm 3.7$ eV b
based on using only potential parameters taken from
elastic scattering at the correct energy.

We have shown that caution should be taken when extracting $S_{17}$ at
center of mass energies approximately greater than 15  MeV due to the fact
that the transfer reaction cannot be considered as peripheral at these
energies, given the angular range usually measured.
 
We have also  shown that the $S_{17}$-factor depends upon
the 2--body  description
of the $^8$B ground state wavefunction if
deformation and/or excitation are included, for optical potentials
that are strongly surface peaked, otherwise the extracted $S_{17}$-factor is
essentially independent of the 2-body description of $^8$B.

Finally we find multistep effects to the  DWBA  to be 
of the order of $8\%$.

We conclude that the ANC method to extract the S-factor from
transfer using the DWBA is accurate to $8\%$  if the 
optical potentials for the incoming and outgoing channels are well defined,
have no strong surface peaked  potentials, and  the reaction is
clearly peripheral.
More measurements at the appropriate low energy regime will help
to further validate the method.

\acknowledgements
This work was supported by Funda\c c\~ao de Ci\^encia
e Tecnologia (Portugal) through grant Praxis PCEX/C/FIS/4/96,
and by EPSRC (U.K.) through grant GR/J95867.
One of the authors, F. Nunes, was supported by FCT grant  BIC 1483.
We would like also to thank Didier Beaumel for providing us with
references to optical potentials.

\begin{table}[tbh]
\caption{Ratio of the asymptotic normalization constant of the w.f to
the direct capture S-factor at 20 KeV.}
\begin{tabular}{cccc}
Model &   $b^{2}$  & $S_{17}^{\rm cap}$  & $b^2/S_{17}^{\rm cap}$ \\
      &(fm$^{-1}$) &      (eV b)         & (eV$^{-1}$ b$^{-1}$ fm$^{-1}$) \\
 \hline 
Kim       & 0.6354 & 23.8 & 0.0267 \\ 
Tombrello & 0.6279 & 23.5 & 0.0267  \\ 
Robertson & 0.6426 & 24.0 & 0.0268  \\ 
Barker    & 0.5946 & 22.4 & 0.0265  \\ 
Esbensen  & 0.4972 & 18.8 & 0.0264  \\ 
\hline
reor1     & 0.6710 & 25.2 & 0.0267  \\ 
reor2     & 0.6967 & 26.0 & 0.0268  \\ 
exc1      & 0.5810 & 22.0 & 0.0264  \\ 
exc2      & 0.6265 & 24.1 & 0.0260  \\ 
 \end{tabular}
\label{as-ratio}
\end{table}


\begin{table}[tbh]
\caption{Comparing the $S_{17}^I$ factors extracted
from transfer data at E$_{cm}$=5.8 MeV with those obtained using
direct capture calculations $S_{17}^{II}$, for different $^8$B g.s models.}
\begin{tabular}{c|cccc|cccc}
      & \multicolumn{4}{c|}{Surface I} &
\multicolumn{4}{c}{Surface II}\\
Model &   ${\cal S}_{exp}$ &   $\chi^2$ &    $S_{17}^I$ &   $S_{17}^{II}$  &
 ${\cal S}_{exp}$  &  $\chi^2$ & $S_{17}^I$   &    $S_{17}^{II}$ \\
      &			   &	        &      (eV b)   &        (eV b)    & 
                   &           &    (eV b)     &      (eV b) \\
 \hline 
Kim       & 1.076 & 0.60 & 26.30 & 25.61   & 0.822 & 0.69 &  20.10  & 19.57  \\ 
Tombrello & 1.048 & 0.69 & 25.31 & 24.63   & 0.832 & 0.69 &  20.10  & 19.55  \\ 
Robertson & 1.017 & 0.72 & 25.14 & 24.41   & 0.813 & 0.70 &  20.10  & 19.52  \\ 
Barker    & 1.138 & 0.58 & 26.03 & 25.50   & 0.881 & 0.66 &  20.15  & 19.74  \\ 
Esbensen  & 1.402 & 0.46 & 26.82 & 26.36   & 1.041 & 0.63 &  19.91  & 19.57  \\ 
\hline
reor1     & 1.005 & 0.54 & 25.94 & 25.32   & 0.652 & 0.73 &  16.83  & 16.43  \\ 
reor2     & 0.968 & 0.64 & 25.93 & 25.16   & 0.731 & 0.72 &  19.59  & 19.01  \\ 
exc1      & 1.170 & 0.54 & 26.15 & 25.75   & 0.749 & 0.67 &  16.73  & 16.47  \\ 
exc2      & 1.099 & 0.60 & 26.47 & 26.47   & 0.862 & 0.65 &  20.77  & 20.77  \\ 
 \end{tabular}
\label{S17-factors}
\end{table}
 


\begin{figure}[t!]
\centerline{
	\parbox[t]{2.5in}{
	\psfig{file=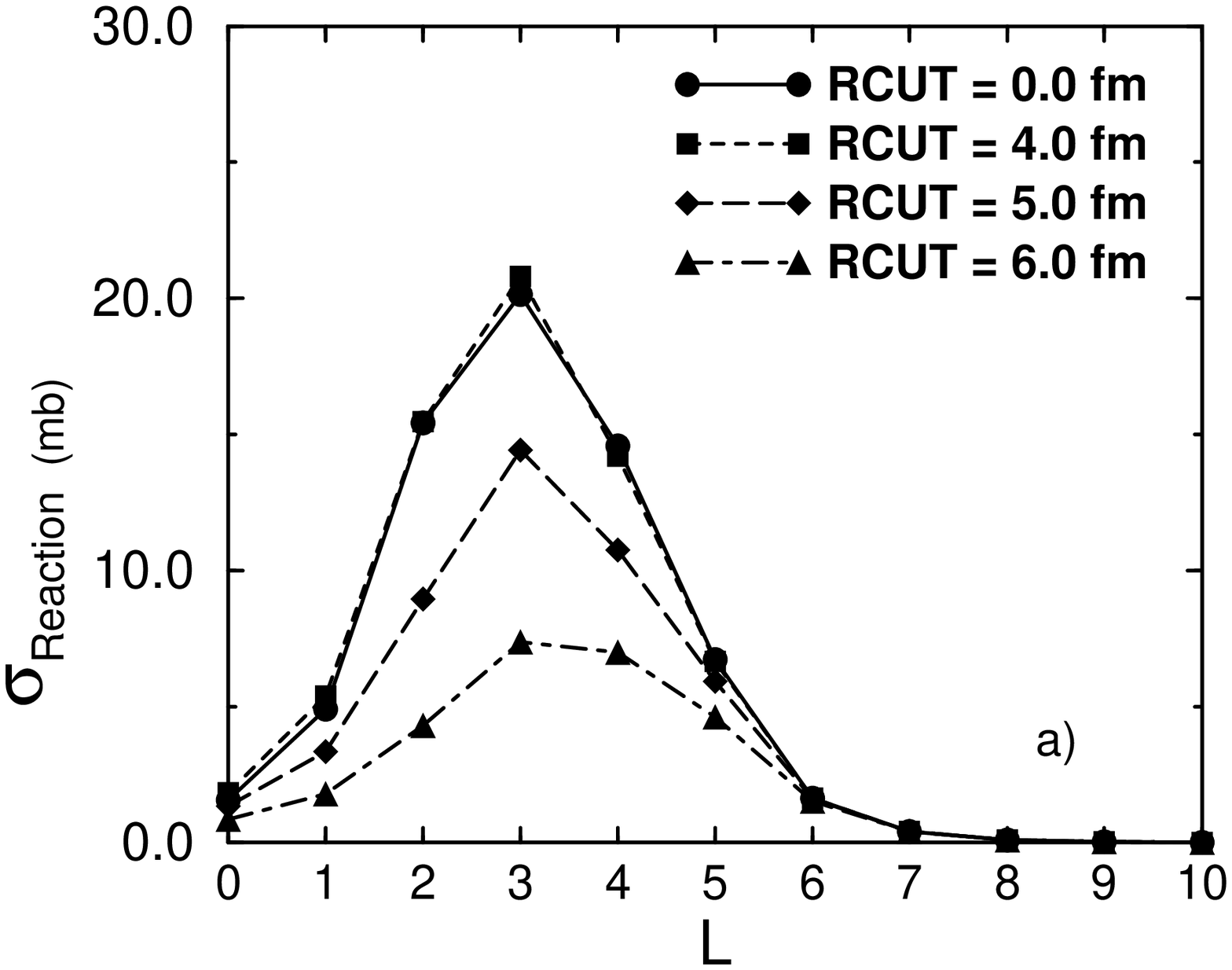,height=2.3in,width=2.3in}}
\hspace{0.3in}
	\parbox[t]{2.5in}{
	\psfig{file=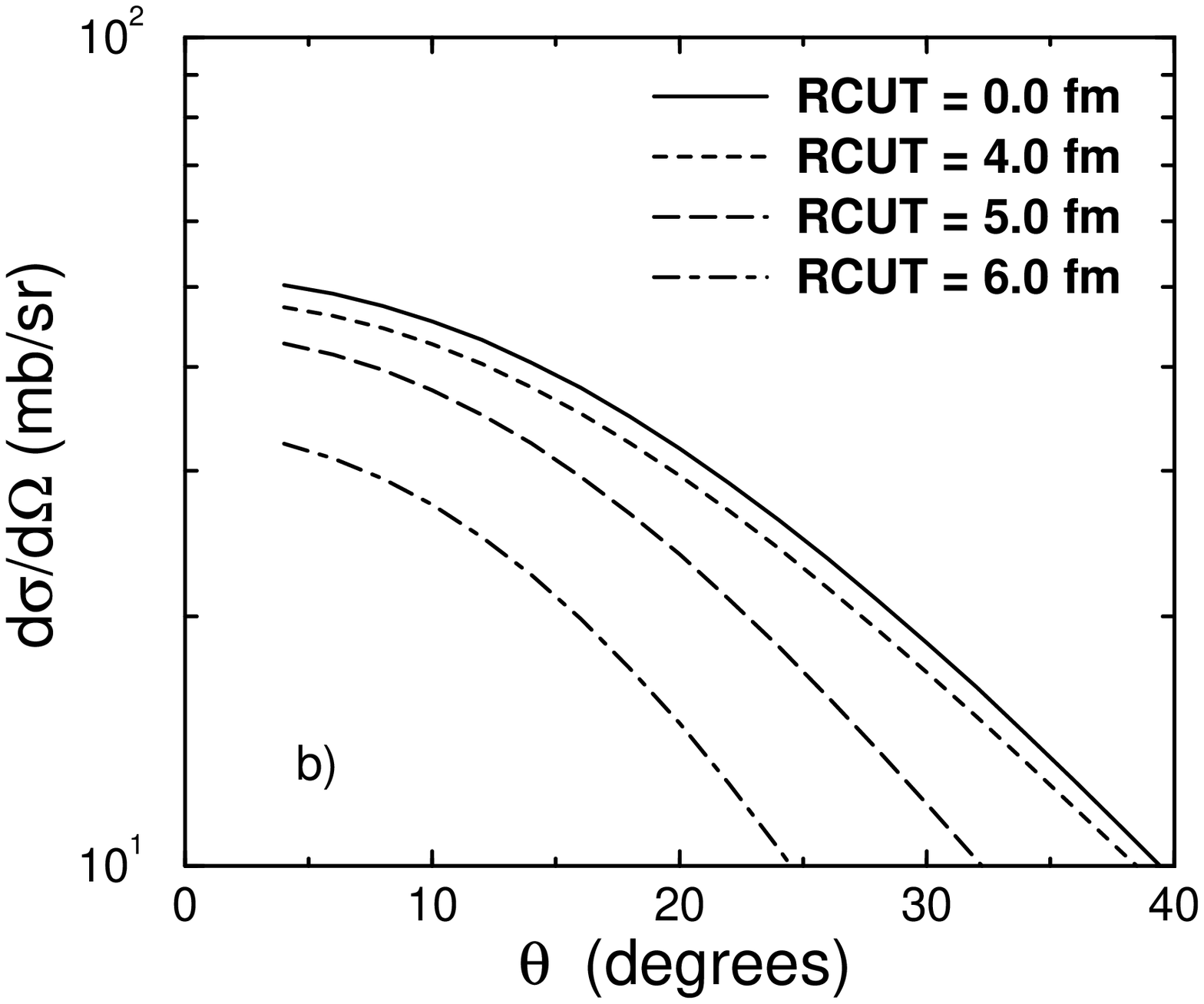,height=2.3in,width=2.3in}}
}
\end{figure}

\begin{figure}[t!]
\centerline{
	\parbox[t]{2.5in}{
	\psfig{file=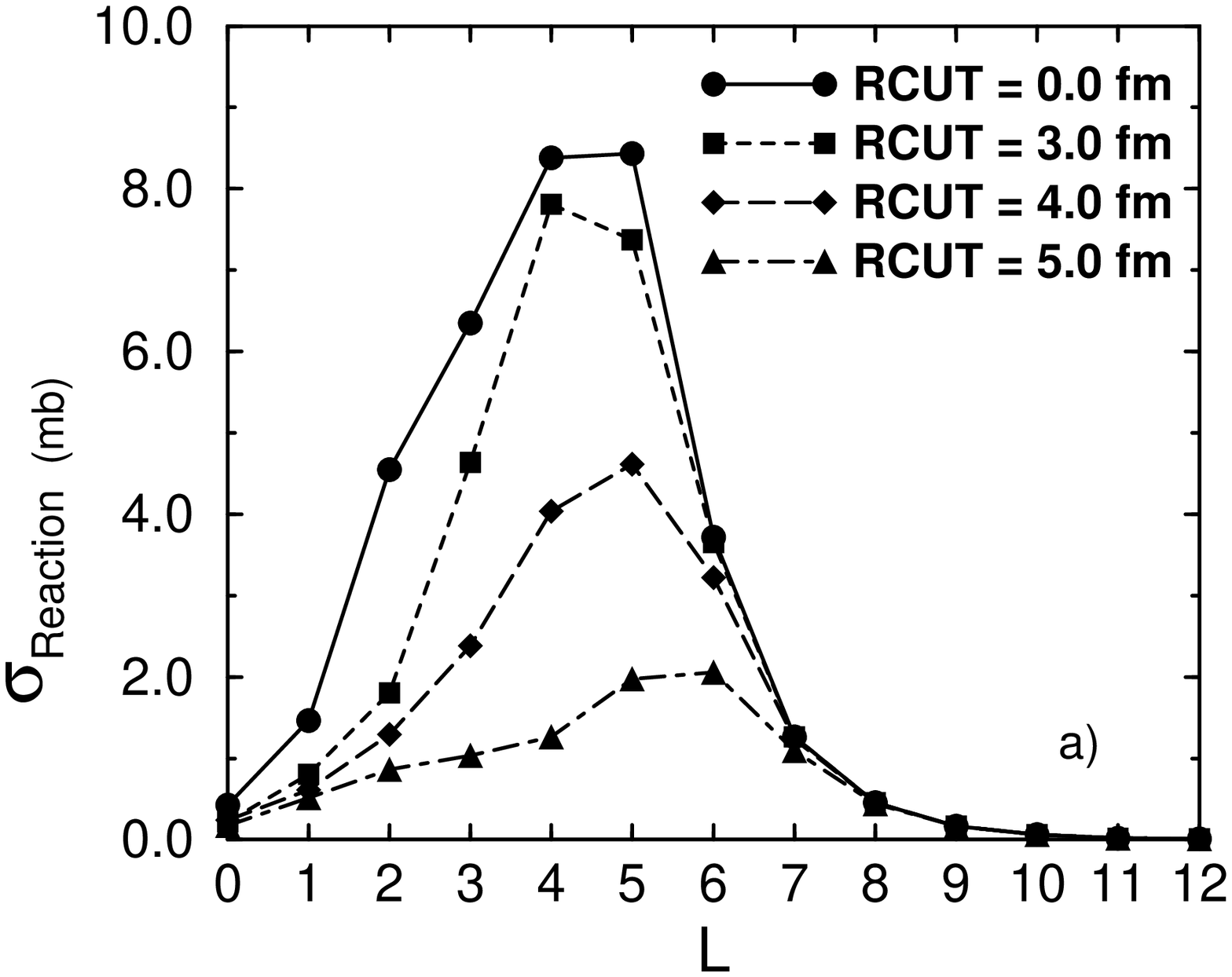,height=2.3in,width=2.3in}}
\hspace{0.3in}
	\parbox[t]{2.5in}{
	\psfig{file=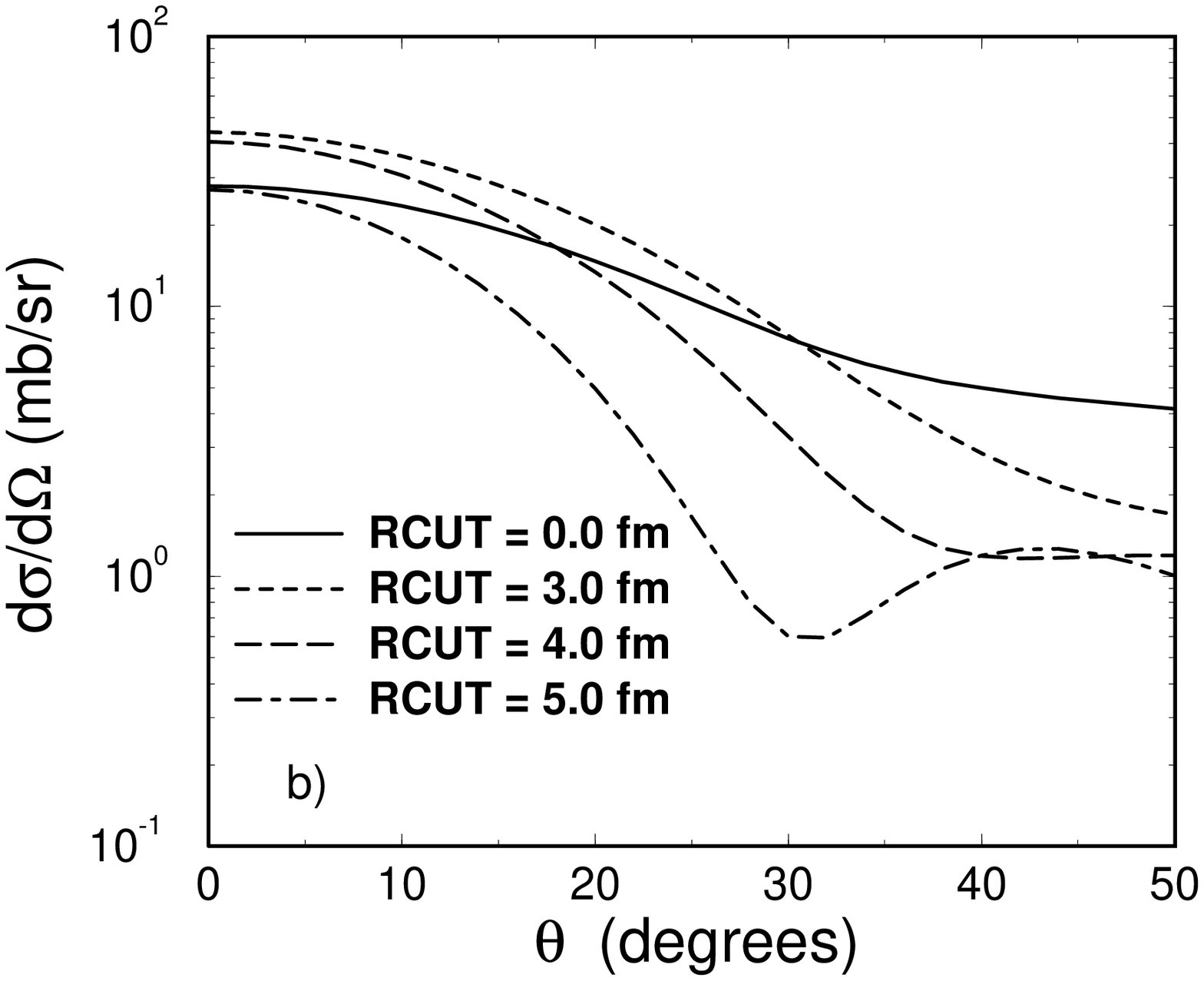,height=2.3in,width=2.3in}}
}
\end{figure}

\begin{figure}[t!]
\centerline{
	\parbox[t]{2.5in}{
	\psfig{file=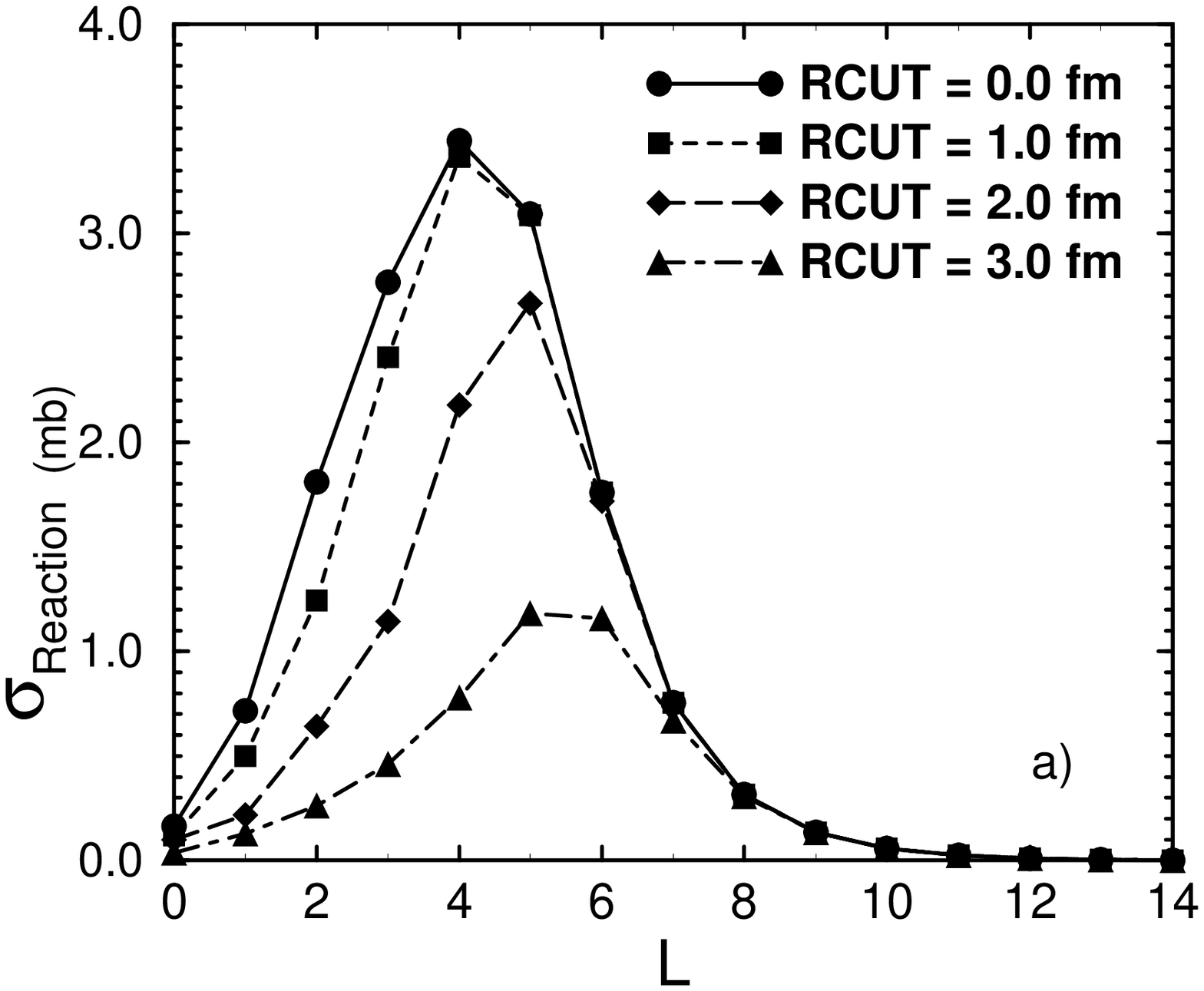,height=2.3in,width=2.3in}}
\hspace{0.3in}
	\parbox[t]{2.5in}{
	\psfig{file=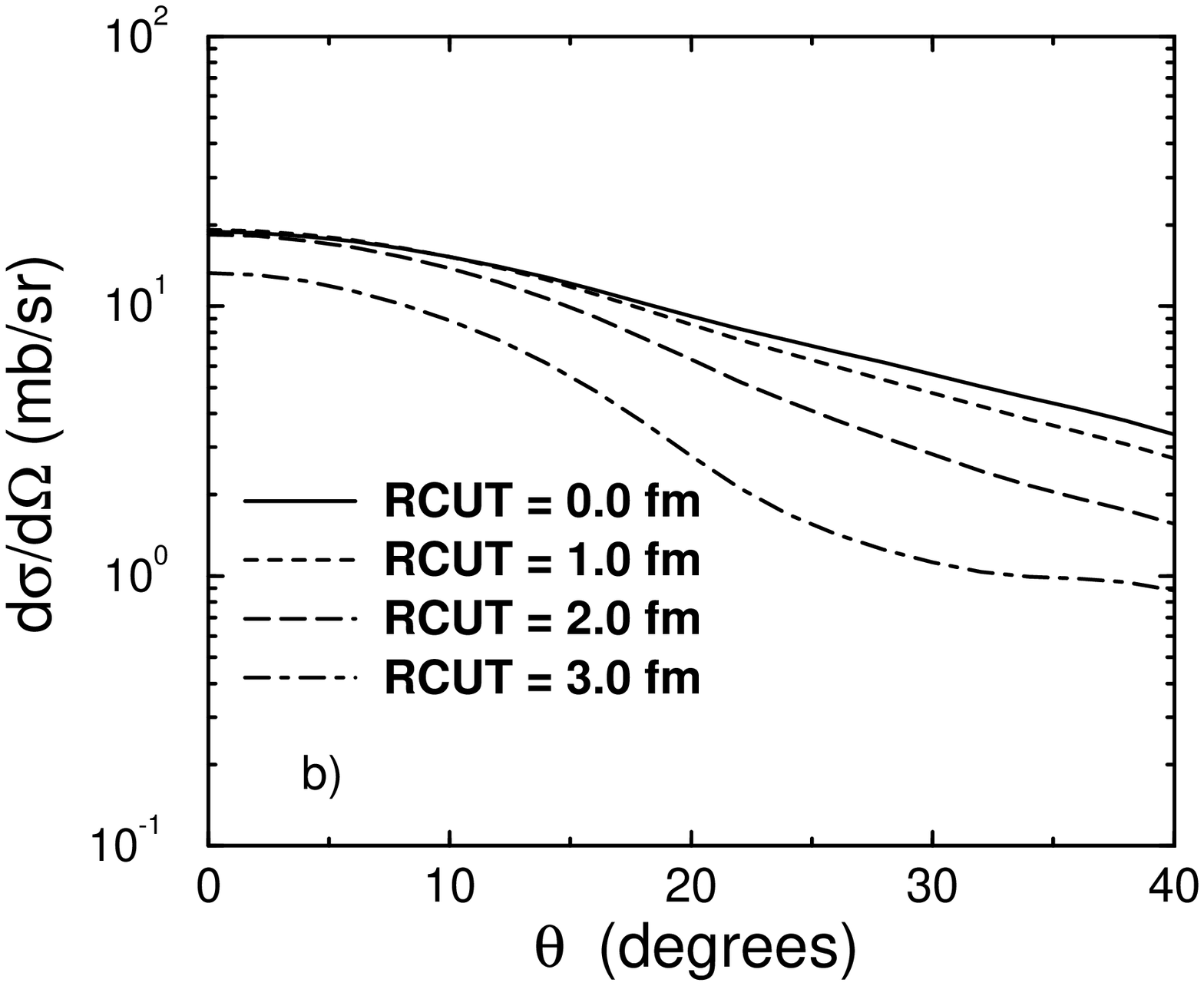,height=2.3in,width=2.3in}}
}
	\caption{Calculated reaction cross section in the  partial wave L,
for the $^7{\rm Be (d,n)}^8{\rm B})$ at 5.8, 15.6 and 38.9 MeV
c.m. energies (a) and corresponding differential cross section (b).}
	\label{fig:xsecl}
\end{figure}


\begin{figure}[t!]
\centerline{
	\parbox[t]{2.5in}{
	\psfig{file=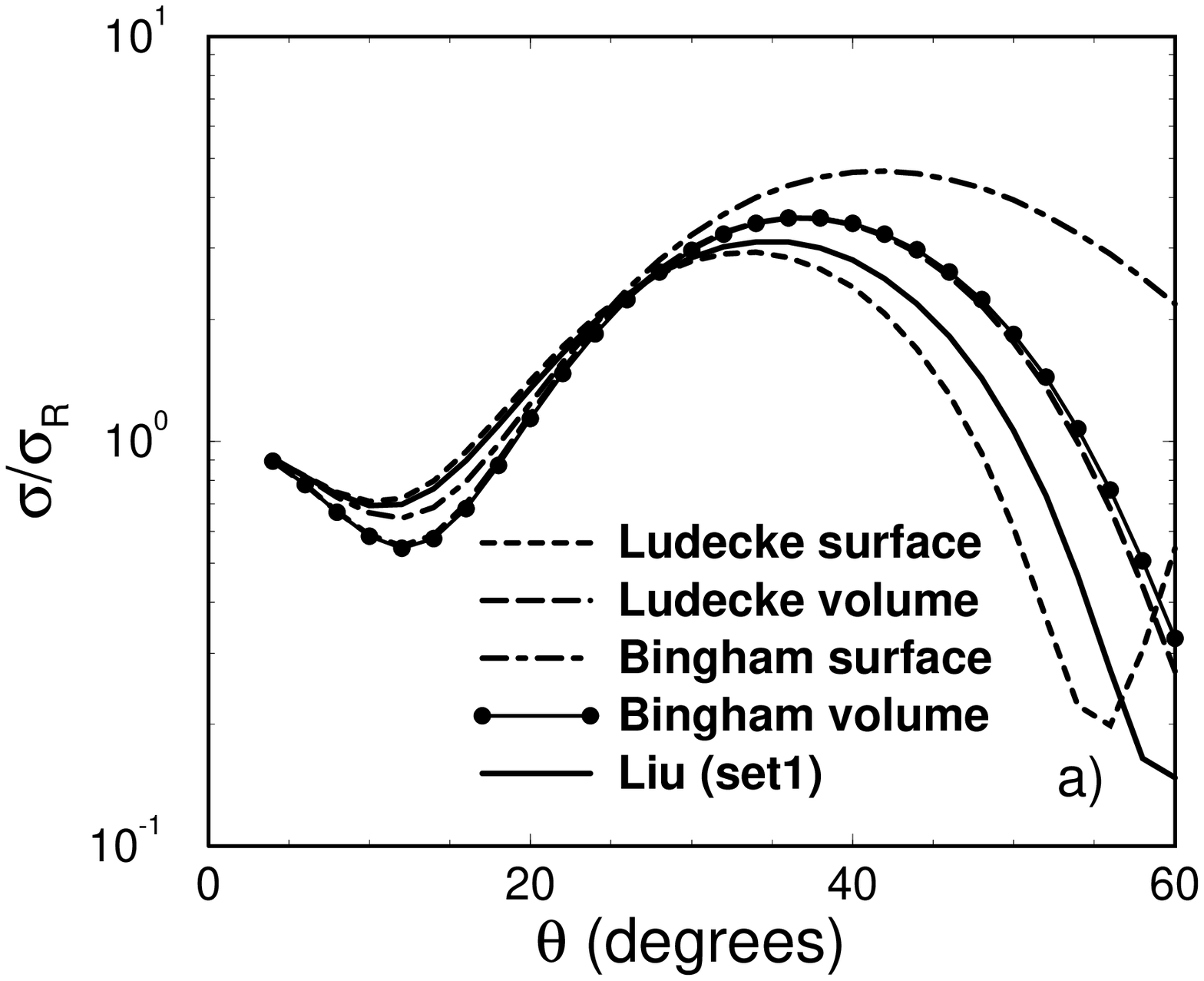,height=2.8in,width=2.8in}}
\hspace{0.3in}
	\parbox[t]{2.5in}{
	\psfig{file=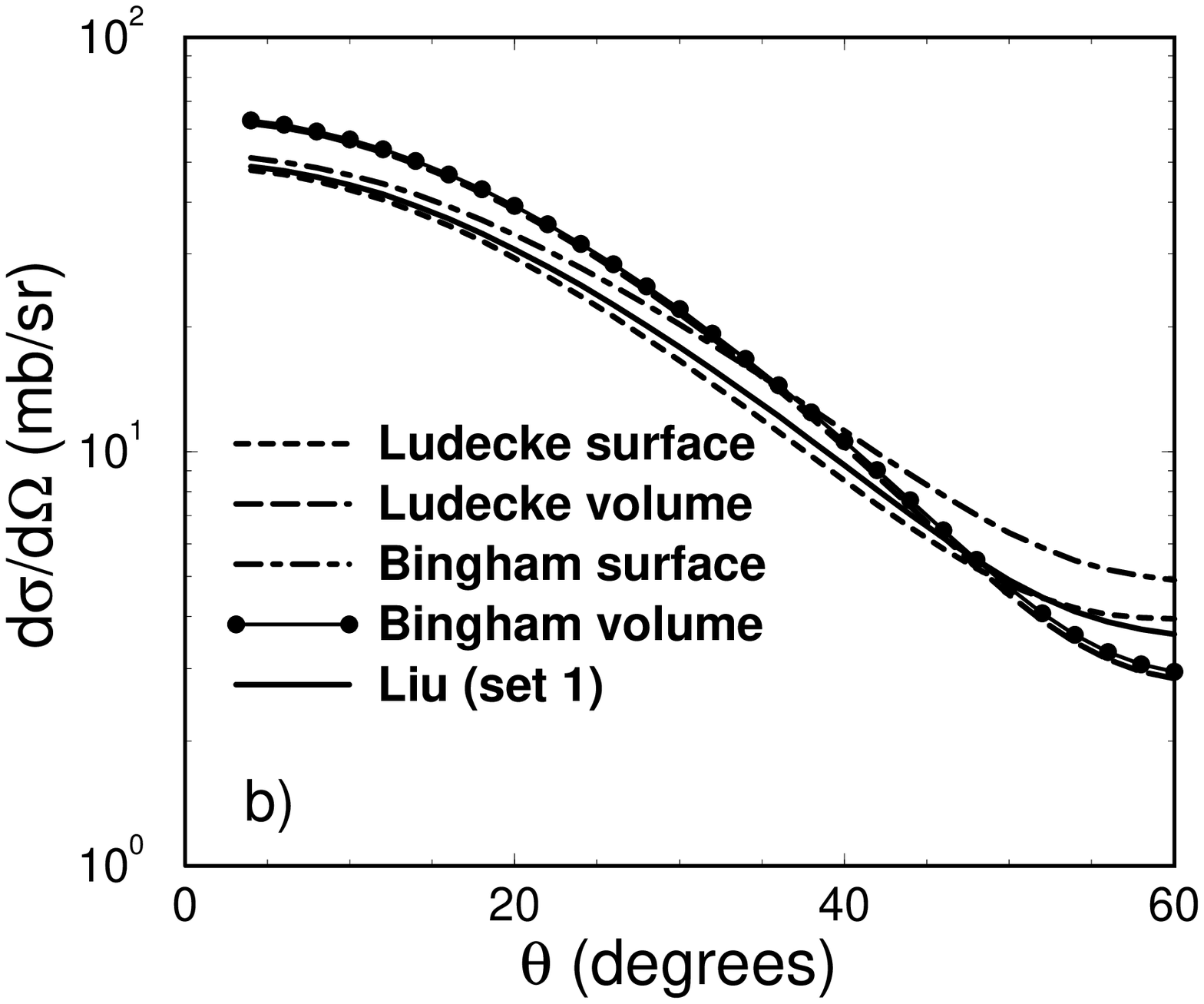,height=2.8in,width=2.8in}}
}
	\caption{The d-$^7$Be elastic scattering at 5.8 MeV
for a set of optical potentials taken from fits to $\simeq$ 9.3  MeV data (a)
and the corresponding transfer differential cross sections (b).}
	\label{fig:elas1}
\end{figure}

\begin{figure}[t!]
\centerline{
	\parbox[t]{2.5in}{
	\psfig{file=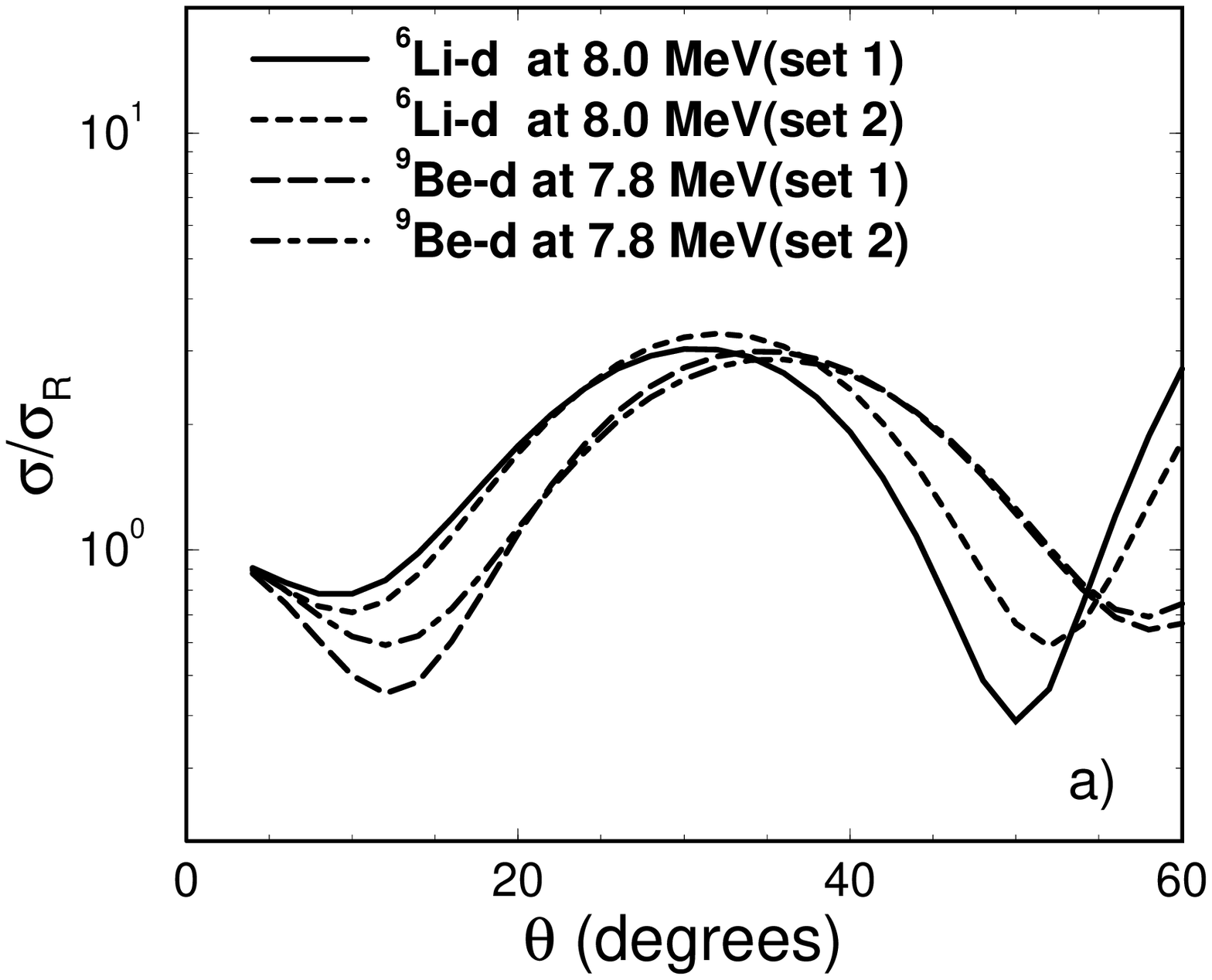,height=2.8in,width=2.8in}}
\hspace{0.3in}
	\parbox[t]{2.5in}{
	\psfig{file=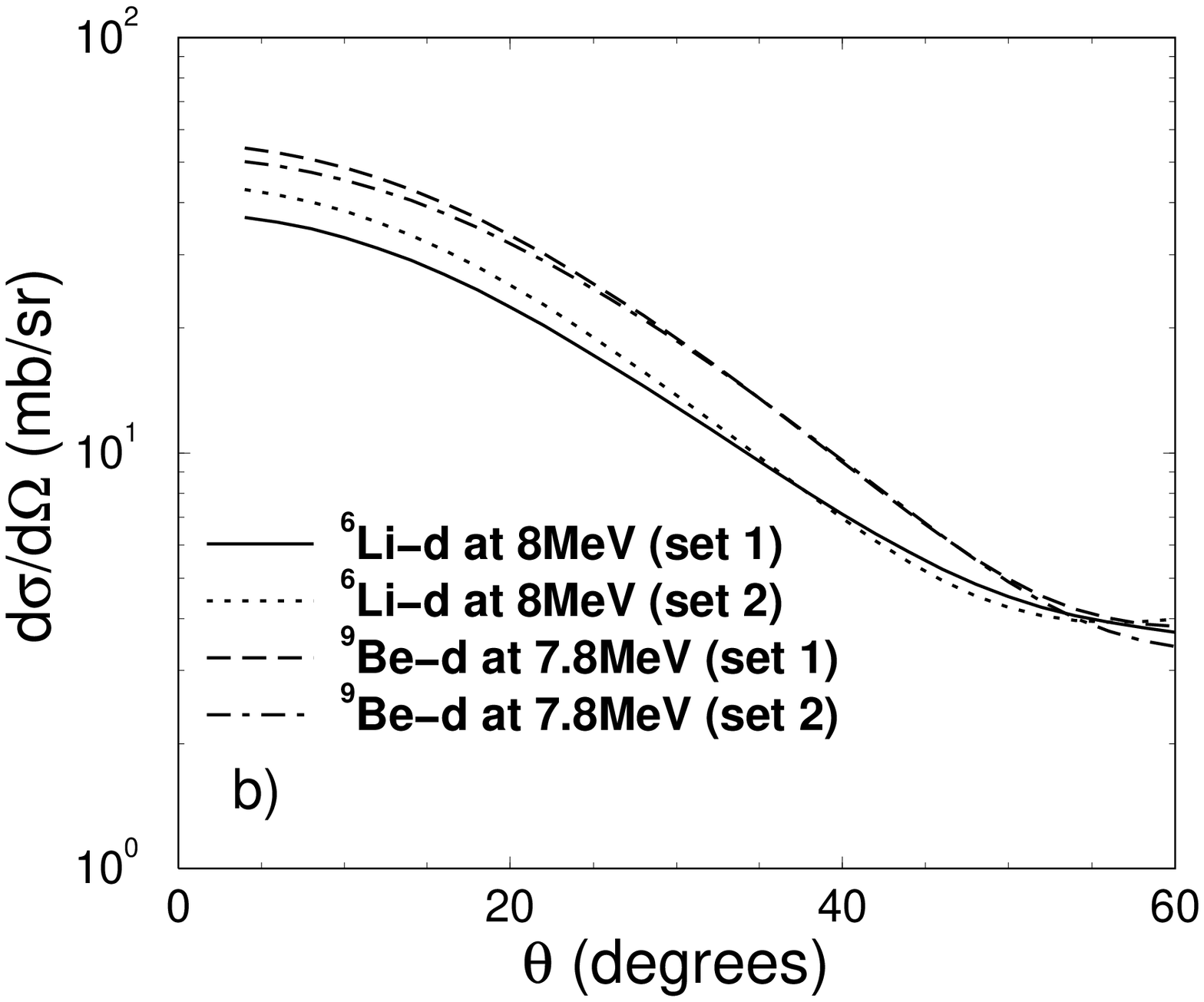,height=2.8in,width=2.8in}}
}
	\caption{The d-$^7$Be elastic scattering at 5.8 MeV
for a set of optical potentials taken at the correct energy but on 
different targets (a) and the corresponding transfer differential
 cross sections (b).}
	\label{fig:elas2}
\end{figure}

\newpage

\begin{figure}[t!]
\centerline{
	\parbox[t]{2.5in}{
	\psfig{file=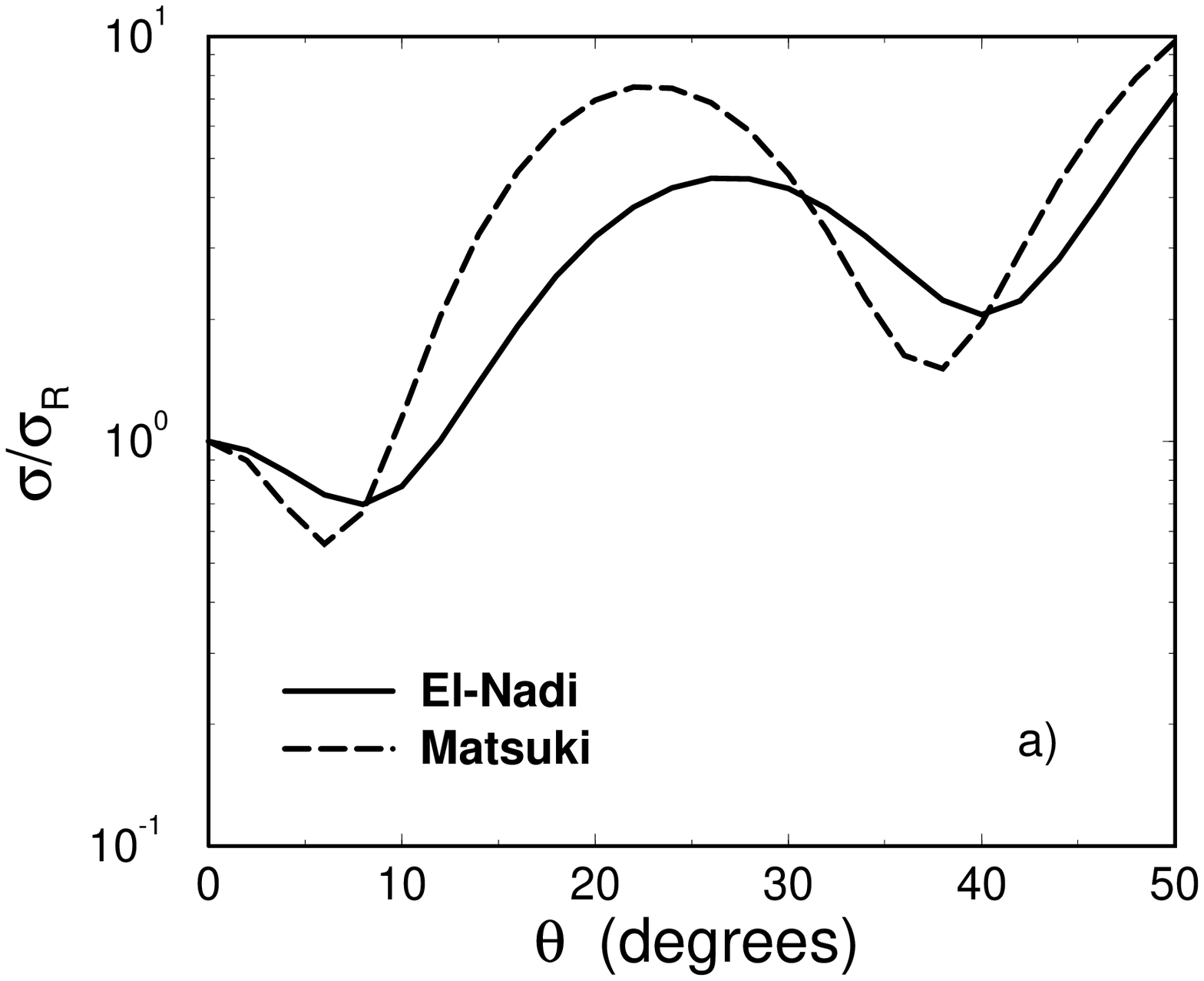,height=2.3in,width=2.3in}}
\hspace{0.3in}
	\parbox[t]{2.5in}{
	\psfig{file=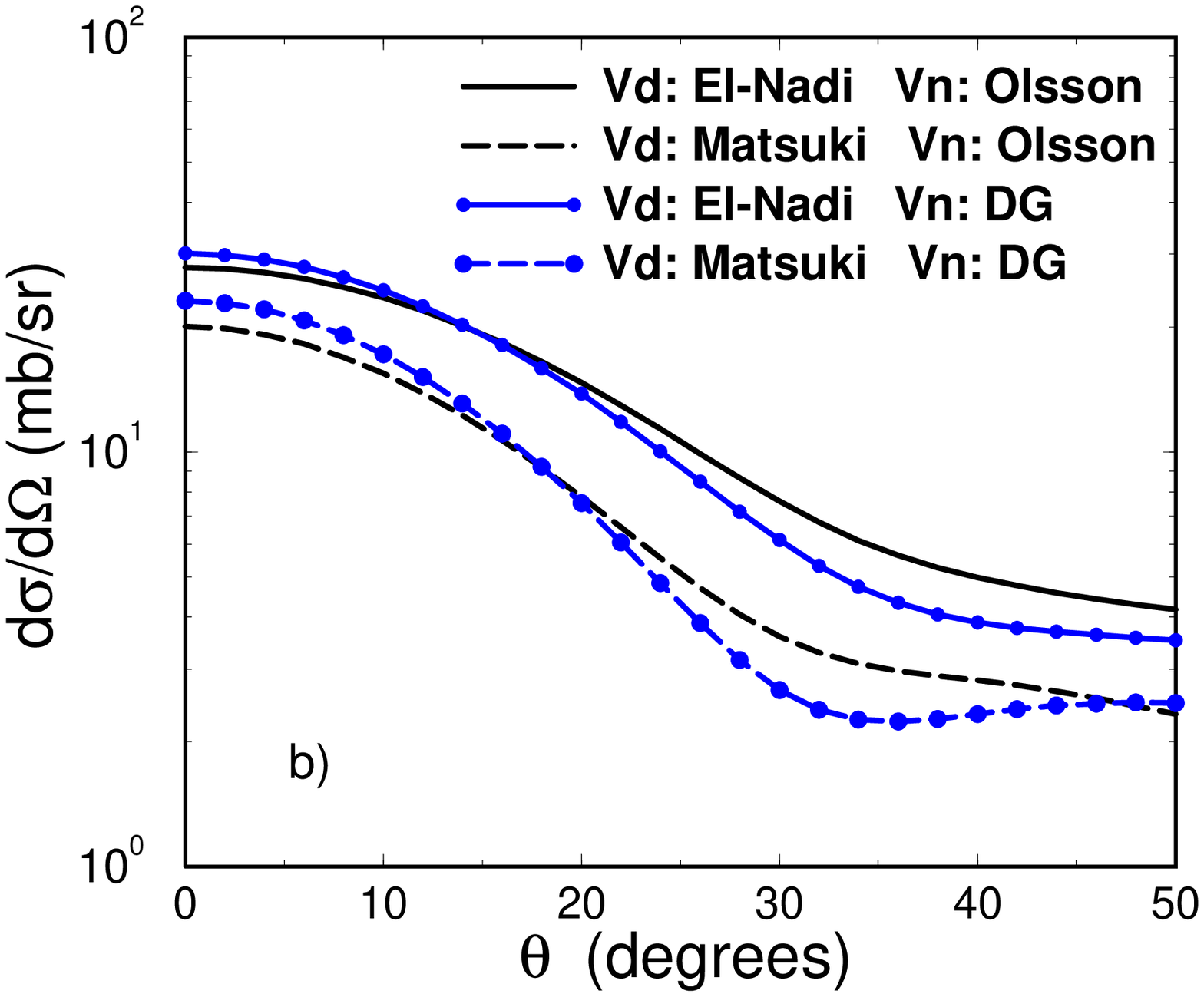,height=2.3in,width=2.3in}}
}
\caption{The d--$^7$Be elastic cross section (a) 
calculated at E$_{\rm cm}$= 15.6 MeV
for two sets of optical potentials (see text), and 
the corresponding transfer differential cross sections (b). }
          \label{fig:Riken}
\end{figure}

\begin{figure}[h!]
\centerline{
	\parbox[t]{2.5in}{
	\psfig{file=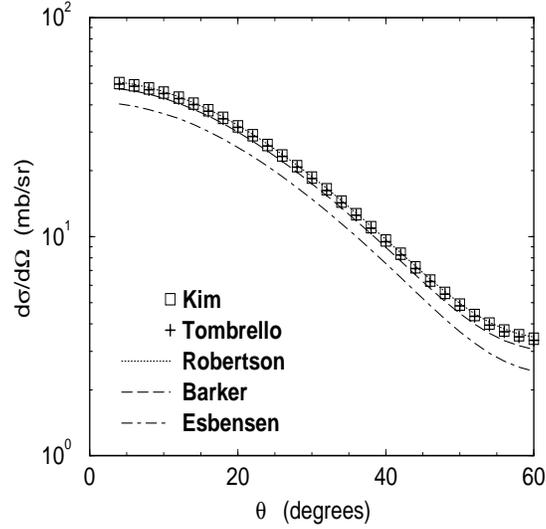,height=2.8in,width=2.8in}}
}
	\caption{Calculated differential cross section for the
$^7$Be(d,n)$^8$B reaction using different models for the $^8$B g.s.
at E$_{\rm cm}$ = 5.8 MeV. }
	\label{fig:B8opt}
\end{figure}

\newpage

\begin{figure}[h!]
\centerline{
	\parbox[t]{2.5in}{
	\psfig{file=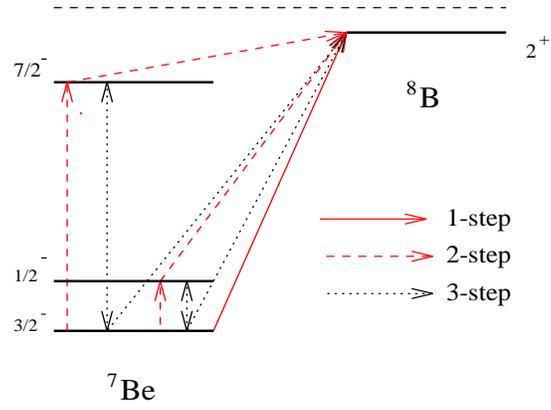,height=2.8in,width=2.8in}}
}
	\caption{The coupling diagram for multistep processes excluding transfer
	back couplings.}
	\label{fig:diag}
\end{figure}


\begin{figure}[t!]
\centerline{
	\parbox[t]{2.5in}{
	\psfig{file=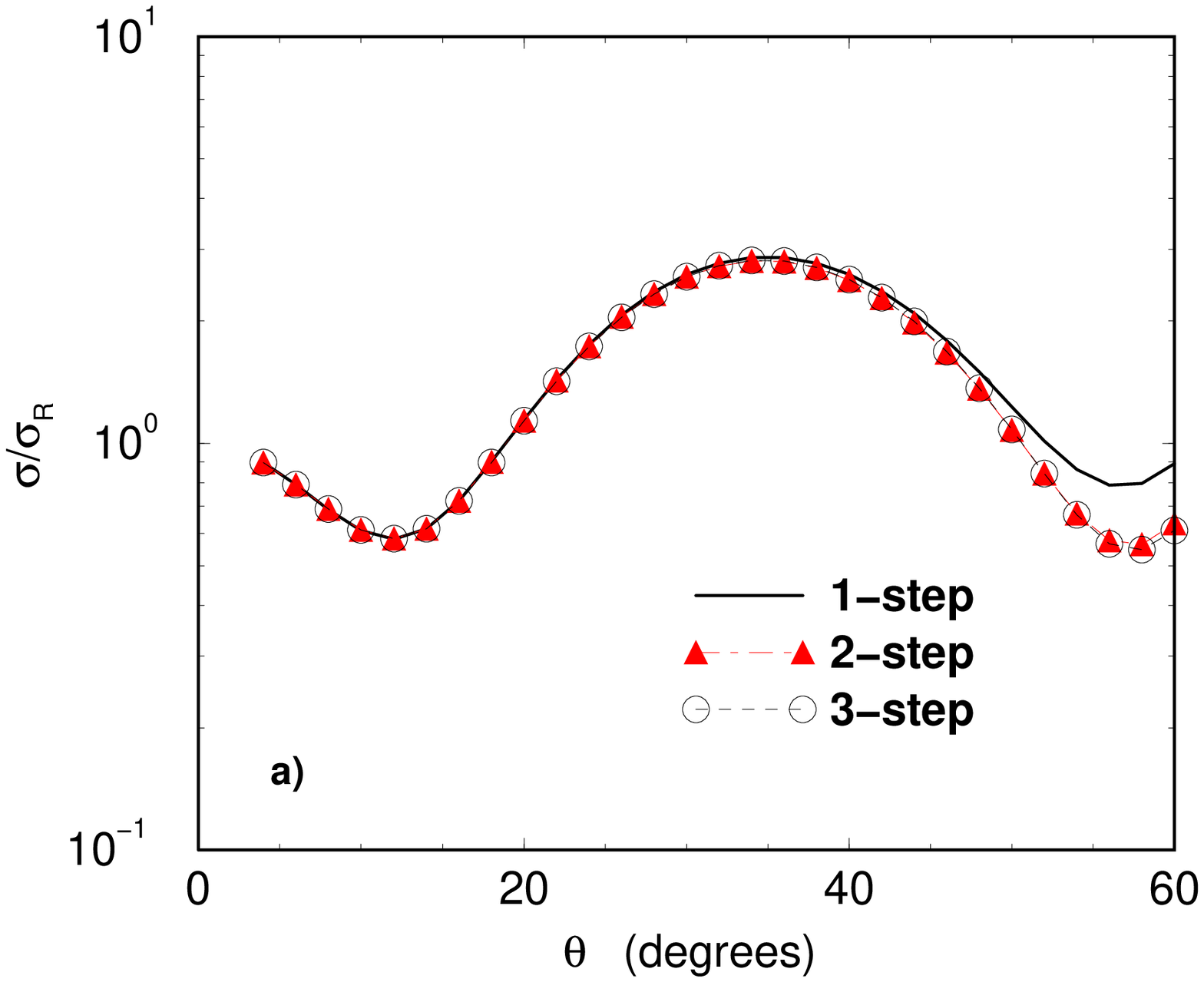,height=2.8in,width=2.8in}}
\hspace{0.3in}
	\parbox[t]{2.5in}{
	\psfig{file=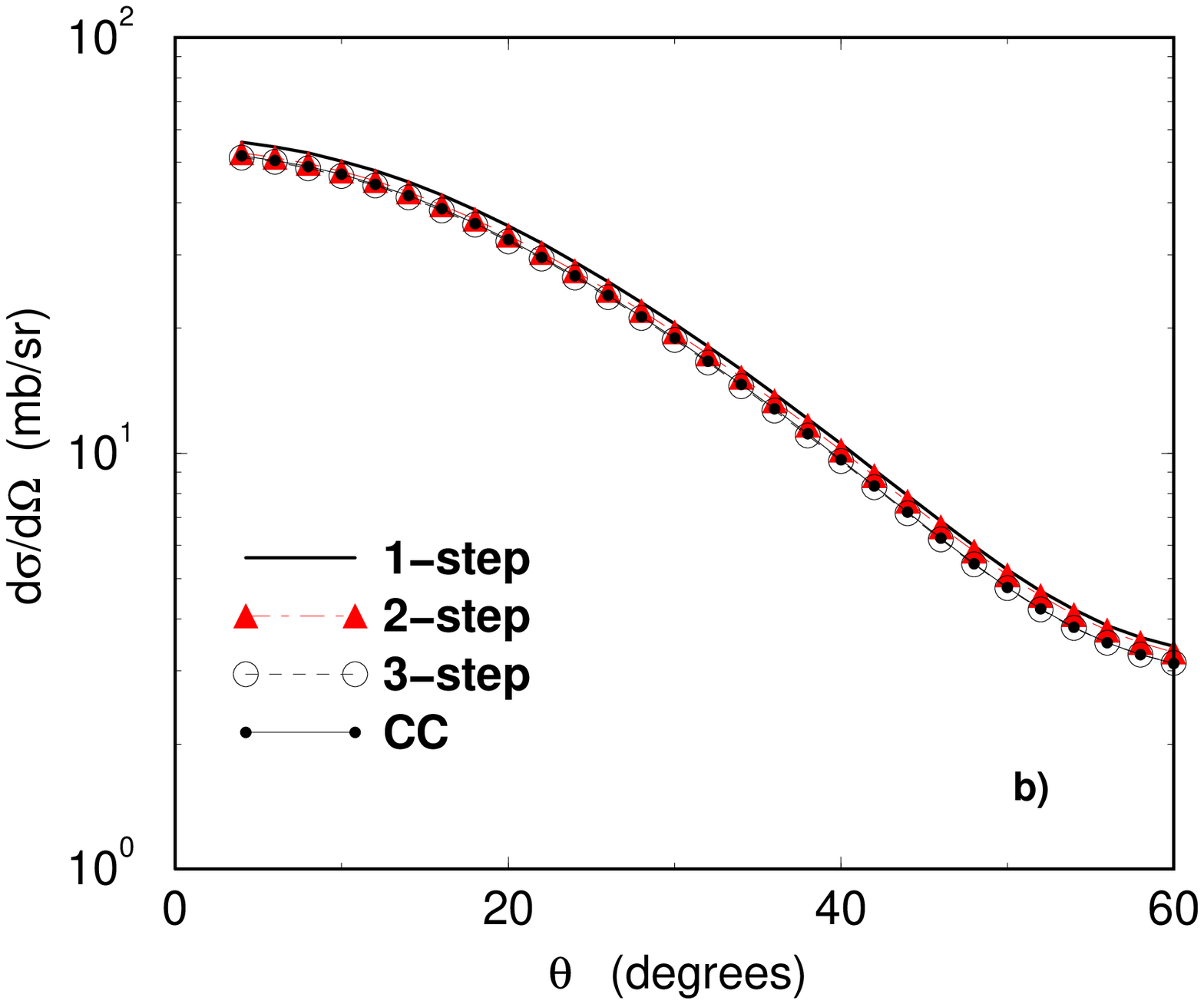,height=2.8in,width=2.8in}}
}
	\caption{a) The elastic scattering (a) and 
 the transfer cross section (b) for 1-step, 2-step, 3-step
and inelastic coupled channel calculations at
 E$_{\rm cm}$ = 5.8 MeV.}
	\label{fig:multi}
\end{figure}

\begin{figure}[t!]
\centerline{
	\parbox[t]{2.5in}{
	\psfig{file=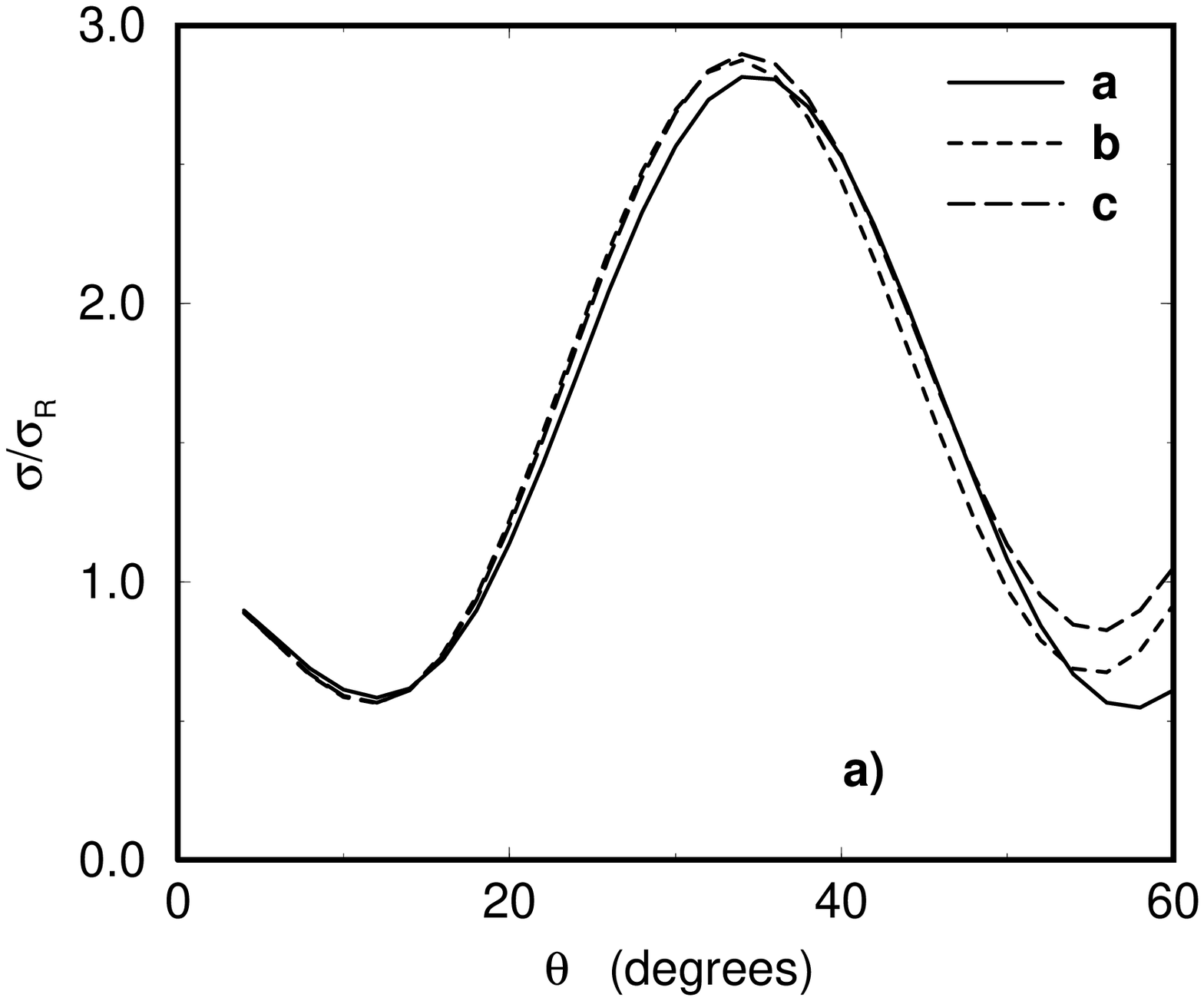,height=2.8in,width=2.8in}}
\hspace{0.3in}
	\parbox[t]{2.5in}{
	\psfig{file=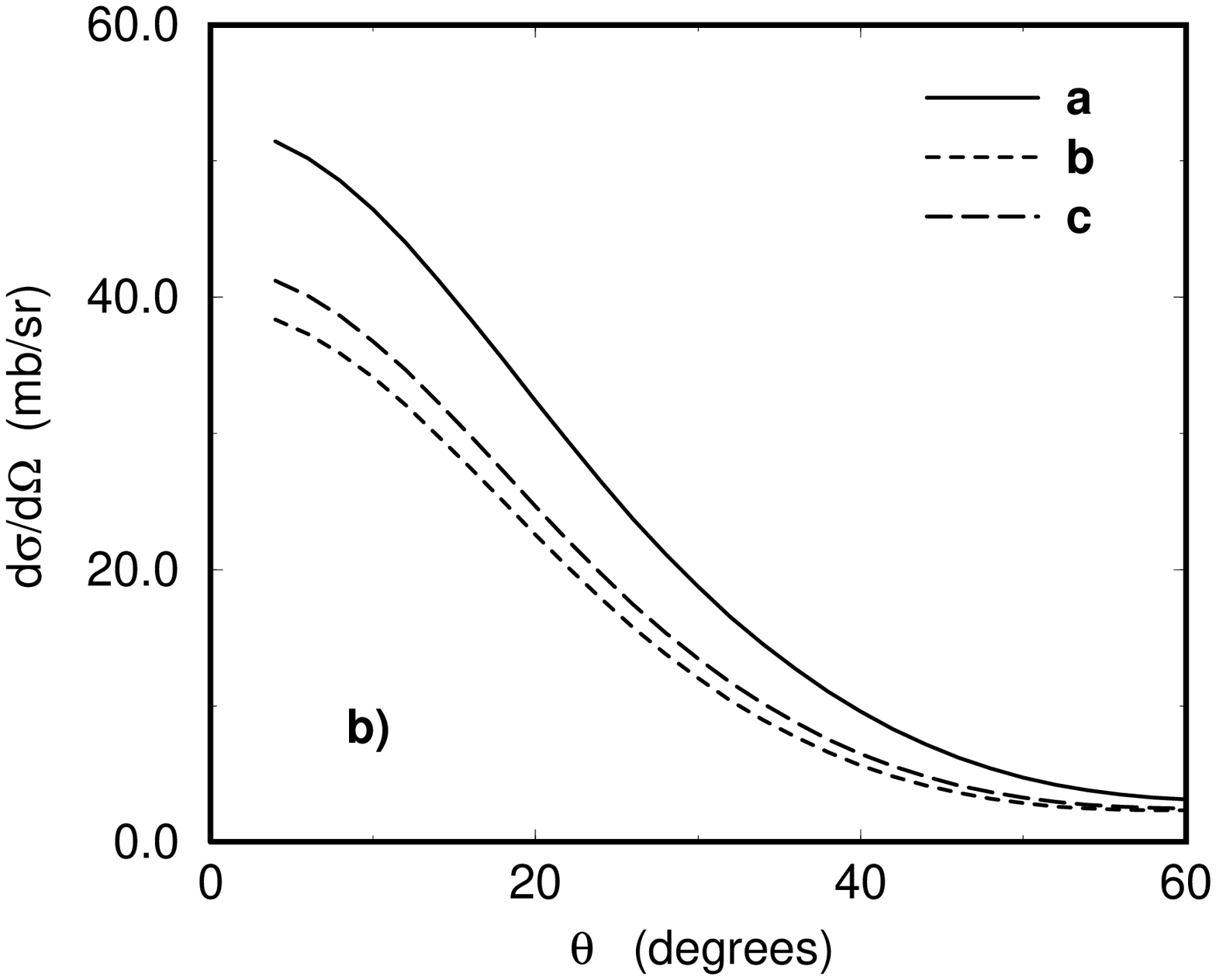,height=2.8in,width=2.8in}}
}
	\caption{Comparing the d-$^7$Be elastic scattering (a)
and transfer reaction (b) when different types of couplings (see text) 
are included at E$_{\rm cm}$ = 5.8 MeV.}
	\label{fig:coup}
\end{figure}

\end{document}